# A GRID OF SYNTHETIC SPECTRA FOR HOT DA WHITE DWARFS AND ITS APPLICATION IN STELLAR POPULATION SYNTHESIS

Ronaldo S. Levenhagen,[1] Marcos P. Diaz,[2] Paula R. T. Coelho,[2] and Ivan Hubeny[3]

[1]*Universidade Federal de São Paulo, Departamento de Física, Rua Prof. Artur Riedel, 275, CEP 09972–270, Diadema, SP, Brazil; ronaldo.levenhagen@gmail.com*

[2]*Universidade de São Paulo, Instituto de Astronomia, Geofísica e Ciências Atmosféricas, Rua do Matão, 1226 São Paulo, SP 05508–900, Brazil*

[3]*Steward Observatory, The University of Arizona, Tucson, AZ 85721, USA*

## ABSTRACT

In this work we present a grid of LTE and non-LTE synthetic spectra of hot DA white dwarfs (WDs). In addition to its usefulness for the determination of fundamental stellar parameters of isolated WDs and in binaries, this grid will be of interest for the construction of theoretical libraries for stellar studies from integrated light. The spectral grid covers both a wide temperature and gravity range, with $17,000\,\text{K} \leq T_\text{eff} \leq 100,000\,\text{K}$ and $7.0 \leq \log g \leq 9.5$. The stellar models are built for pure hydrogen and the spectra cover a wavelength range from 900 Å to 2.5 $\mu$m. Additionally, we derive synthetic HST/ACS, HST/WFC3, Bessel UBVRI and SDSS magnitudes. The grid was also used to model integrated spectral energy distributions of simple stellar populations and our modeling suggests that DAs might be detectable in ultraviolet bands for populations older than $\sim 8\,\text{Gyr}$.

*Keywords:* techniques: spectroscopic — stars: atmospheres — line: profiles — stars: white dwarfs — stars: fundamental parameters — globular clusters: general

## 1. INTRODUCTION

White dwarfs (WDs) are the remnants of about 97% of the stars scattered along the HR diagram with main-sequence (MS) masses ranging approximately from 0.07 to 10.5 $M_\odot$, depending on metallicity and binarity (Fontaine et al. 2001; Doherty et al. 2015; Kepler et al. 2015). A huge effort in the last years has been devoted to determine the cooling rates and ages of WDs.

As compact objects, they present large densities (around $10^9 \, \mathrm{kg \cdot m^{-3}}$), low luminosities ($10^{-2}$ - $10^{-5} \, L_\odot$), small radii (of the order of an ordinary planet), and large surface gravities ($\log g \geq 7$ c.g.s. units). They are mainly divided into DA, DO, DB, and DC, depending on the chemical composition of the atmosphere. DA WDs exhibit pure hydrogen spectra and occur at all temperatures in the cooling sequence (Koester 2002). DO WDs are usually hotter than 45,000 K; DB stars range from 11,000 K and 30,000 K; and DC dwarfs are cooler than 11,000 K. The chemical composition of the atmosphere is defined mainly by their high surface gravity values, as the chemical elements are stratified according to their atomic weights (Dreizler & Wolff 1999).

There are currently few synthetic DA spectral grids available for general application in the literature, and they are designed for specific temperature and wavelength ranges. Bergeron et al. (1991) proposed a grid of model atmospheres and synthetic spectra for cool DA WDs, covering the ranges $5,000 \, \mathrm{K} \leq T_\mathrm{eff} \leq 12,000 \, \mathrm{K}$ and $7.5 \leq \log g \leq 8.5$.

In the present work we have built a homogeneous grid of model spectra for DA WDs in LTE and NLTE. The synthetic spectra cover the UV at 900 Å up to the IR at 2.5 $\mu$m. They are computed between $17,000 \, \mathrm{K} \leq T_\mathrm{eff} \leq 100,000 \, \mathrm{K}$ in steps of 1,000 K and $7.0 \leq \log g \leq 9.5$ in steps of 0.1 dex, taking into account quasi-molecular opacities (Allard et al. 2009) in cooler DA models with $17,000 \, \mathrm{K} \leq T_\mathrm{eff} \leq 30,000 \, \mathrm{K}$.

Additionally, we derive synthetic integrated magnitudes for Bessel UBVRI, Sloan Digital Sky Survey (SDSS), HST/ACS, and HST/WFC3 photometric systems, which can be useful to find solutions for a wide range of problems involving precision photometry of WDs in binaries, multiple systems, and clusters. Model magnitudes may be also used to define magnitude transformations between those systems for WDs.

We also employed this grid in an application to quantify the effect of these stars in the integrated light of populations using stellar population modeling techniques. It is known that in visual bands, the effect of the WD population is minimal in integrated light (e.g. Bruzual A. & Charlot 1993; Bruzual & Charlot 2003). Nevertheless, it is reasonable to assume that they might have a measurable impact in UV colors, which we quantify in Section 5.

## 2. ATMOSPHERE STRUCTURE AND SPECTRAL SYNTHESIS

All the synthetic spectra of DA WDs presented in this paper were generated using the spectrum synthesis code SYNSPEC (Hubeny & Lanz 2011), version 50. Corresponding model atmospheres were computed by the code TLUSTY (Hubeny 1988; Hubeny & Lanz 1995), version 205. For a description and user's manual for the newest versions, see Hubeny & Lanz (2017).

Model atmospheres were constructed under the standard assumptions of a plane-parallel, horizontally homogeneous (i.e., 1-dimensional) atmosphere in hydrostatic and radiative equilibrium. Since we consider DA WDs, a pure hydrogen chemical composition was assumed. Convection was neglected in the atmosphere, although a thin convectively unstable zone may be present at low depths for the lower effective temperatures in our grid (Tremblay & Bergeron 2008). However, significant convective fluxes should occur only for $T_\mathrm{eff} \lesssim 14,000 \, \mathrm{K}$ (depending on gravity) (Tremblay et al. 2015), therefore small or negligible changes to the temperature profile due to convection are expected for our models, which have $T_\mathrm{eff} \geq 17,000 \, \mathrm{K}$.

For models with $T_\mathrm{eff} < 34,000$ K we are assuming LTE, while for hotter models, we computed NLTE models, allowing for departures from LTE for the first 16 levels of hydrogen. For all models, we consider a level dissolution directly in the evaluation of line profiles using the occupation probability formalism of Hummer & Mihalas (1988), adapted for NLTE models by Hubeny et al. (1994). The cumulative opacity of transitions from the first three levels of hydrogen to higher, dissolved levels, is treated using the concept of a pseudo-continuum, after Hubeny et al. (1994). Line broadening was treated using the improved unified (impact plus quasi-static) theory of Tremblay & Bergeron (2009) that takes into account the effects of level dissolution directly in the evaluation of line profiles.

For $17,000 \, \mathrm{K} < T_\mathrm{eff} < 30,000 \, \mathrm{K}$, we first computed simple models not including the quasi-molecular satellites of Ly$\alpha$, $\beta$, and $\gamma$ that arise due to close H–H+ collisions (Allard et al. 2009), and then final models were calculated with these satellites.

All the synthetic spectra were computed to cover the wavelength range from 900 Å to 2.5 $\mu$m. The microturbulent velocity was set to 0 km s$^{-1}$ for all models. The resulting spectrum was subsequently processed with the code ROTIN (Hubeny & Lanz 2017), with the rotational velocity as well as the instrumental resolution set to zero; the code thus only resampled the original synthetic spectrum. All database spectra are provided in Eddington flux units at the WD surface (in erg cm$^{-2}$ s$^{-1}$ Å$^{-1}$).

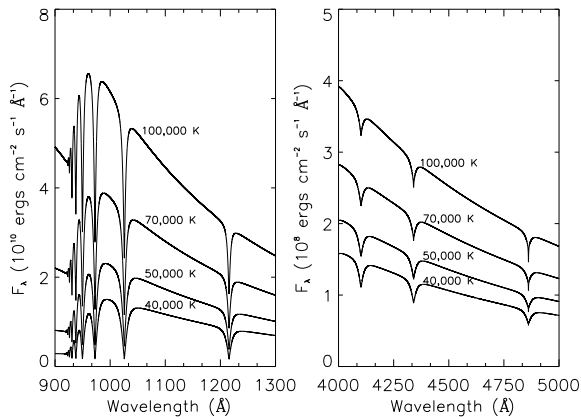

**Figure 1.** Theoretical non-LTE spectra of a pure hydrogen DA white dwarf with $T_{\rm eff}$ between 40,000 K and 100,000 K and $\log g = 7.0$ without quasi-molecular H-H+ absorption.

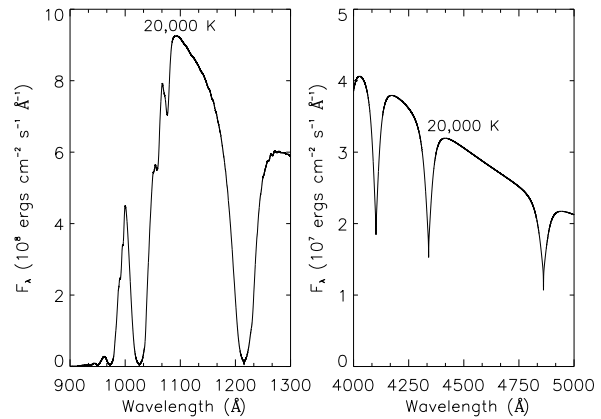

**Figure 2.** Theoretical LTE spectrum of a pure hydrogen DA white dwarf with $T_{\rm eff} = 20,000$ K and $\log g = 7.0$ taking into account the satellites of quasi-molecular H-H+.

A subset of the resulting grid is shown in Fig. 1. A total of 2,183 synthetic spectra have been computed (84 temperatures and 26 gravities). Among them, 1,716 assume NLTE. The final data are made available in a digital form and a specific or extended spectral synthesis can be provided by the authors upon request. The grid repository can be accessed via the Vizier service and at `http://specmodels.iag.usp.br`. An example of the effect of quasi-molecular opacities is shown in Fig. 2. An Inglis–Teller relation derived from the model structure and spectra is shown in Fig. 3, aiming to illustrate the behavior of the density near the photosphere as a function of the highest observable term $N_{\rm max}$ in the series.

## 3. SYNTHETIC MAGNITUDES

In this work we have calculated model atmosphere magnitudes for 90 photometric bands comprising the Bessel UBVRI, SDSS, HST/ACS, and HST/WFC3 photometric systems. The filter response functions ($S$) of these photometric systems available in the Filter Profile Service at the Spanish Virtual Observatory[1] (SVO) were used to evaluate the synthetic integrated fluxes. Bessel UBVRI, HST/ACS (VEGAmag,[2] Sirianni et al.

[1] `http://svo.cab.inta-csic.es/main/index.php`
[2] The zero-point transformation to the STmag system can be found at `acszeropoints.stsci.edu`

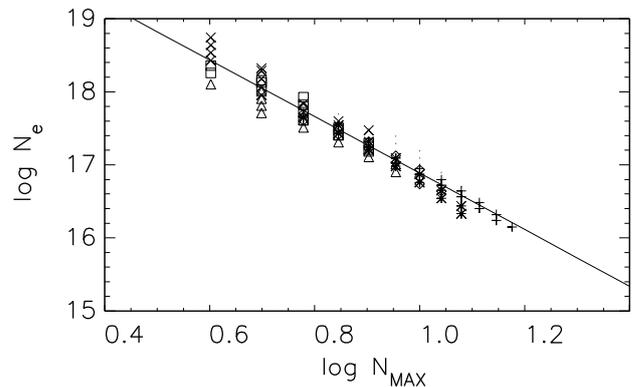

**Figure 3.** Inglis–Teller diagram for models in the range 17,000 K $\leq T_{\rm eff} \leq$ 100,000 K. The best fitting yields $\log N_e = \alpha \log N_{\rm max} + \beta$ with $\alpha = -3.87 \pm 0.09$ and $\beta = 20.76 \pm 0.09$. It is worth noting that the electronic density reference is set at the atmospheric layer where $\tau_{\rm Rosseland} \simeq 1$.

(2005)), and HST/WFC3 (VEGAmag[3]) magnitudes were calculated in the Vega system while the SDSS synthetic mags are bandpass AB magnitudes. For the Vega systems the magnitudes and zero-points were found from the band-averaged flux density $F(s)$ as follows:

$$F(s) = \frac{\int S(\lambda) f(\lambda) \lambda d\lambda}{\int S(\lambda) \lambda d\lambda} \quad (1)$$

where $S(\lambda)$ is the filter response function and $f(\lambda)$ is the physical flux. The emerging model flux output of

[3] The zero-point transformation to the STmag system can be found at `www.stsci.edu/hst/wfc3/phot_zp_lbn`

ROTIN is expressed as the Eddington flux H($\lambda$, $T_{\rm eff}$, log $g$) at the stellar surface. It is converted to physical flux $f(\lambda)$ at a distance of 10 parsecs, assuming a simple zero-temperature, non-relativistic mass-radius relation for carbon WDs (Hamada & Salpeter 1961):

$$f(\lambda) = 4\pi \left( \frac{R^2}{D^2} {\rm H}(\lambda, {\rm T}_{\rm eff}, \log g) \right) \quad (2)$$

where $\pi \frac{R^2}{D^2}$ stands for the solid angle, or the ratio between the radius of the star $R$ and its distance $D$. All magnitudes are scaled for a geometric dilution by $D = 10$ pc.

The integration is performed over the filter bandpass and the bandpass-averaged flux density $F(s)$ (in erg cm$^{-2}$ s$^{-1}$ Å$^{-1}$) is used to compute synthetic absolute magnitudes (Figure 4) for each photometric band:

$$M_s = -2.5 \log F_s + ZP \quad (3)$$

where $ZP$ are the corresponding Vega system zero-points, with the Vega magnitude being identically zero in ACS and WFC3 systems and +0.03 in Bessel UBVRI.

The synthetic SDSS system absolute magnitudes were calculated using the definition of broadband AB magnitudes by Fukugita et al. (1996), with a zero-point bandpass-averaged flux of 3631 Jy. The sensitivity functions $S(\nu)$ defining the survey photometric system for a point source were obtained at the SDSS DR7 website[4]. Outside atmosphere SDSS magnitudes are calculated from flux densities (eq. 2) in $f(\nu)$ units. It is worth pointing out that the SDSS "$u$" band magnitudes were previously reported as systematically different from their corresponding synthetic AB magnitudes by 0.04 mag (i.e. $u_{\rm AB} = u_{\rm SDSS} - 0.04$) (Holberg & Bergeron 2006). No correction was applied to the synthetic SDSS "$u$" magnitudes calculated here.

## 4. GRID BENCHTEST

In order to test the grid of synthetic spectra we chose to compare stellar parameters inferred for DA standards. Instead of individual line profile fitting, we have chosen a global flux fitting (including the continuum) of a well-known standard flux star. Model fitting to the observed spectrum of the standard WD GD71 (J2000 $\alpha = 05^{\rm h}52^{\rm m}27.5^{\rm s}$, $\delta = +15^{\rm o}53'17''$, WD DA1.5 C, $B - V = -0.249$, $V = 13.032$) is performed using a downhill simplex algorithm (AMOEBA) (Nelder & Mead 1965). A first guess of the stellar distance is given by the

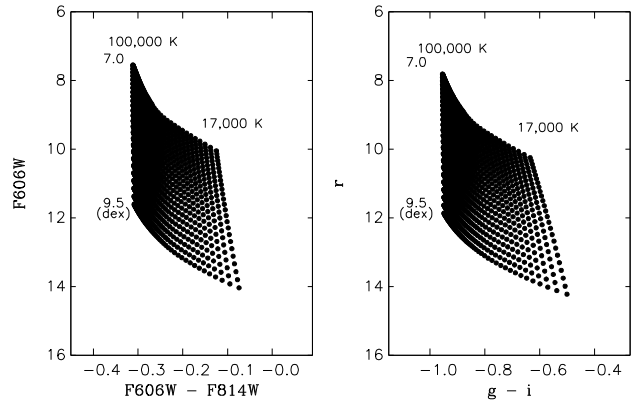

**Figure 4.** (Left panel) Sample synthetic color-magnitude diagram for DA WDs in the HST/WFC3 F606W and F814W bands for all models in the ranges $17,000\,{\rm K} \leq T_{\rm eff} \leq 100,000\,{\rm K}$ and $7.0 \leq \log g \leq 9.5$. (Right panel) Sample synthetic color-magnitude for the same models in the AB SLOAN SDSS bands.

Sion & Liebert (1977) absolute magnitude empirical formula:

$$M_v = 11.246(B - V + 1)^{0.6} - 0.045 \quad (4)$$

The GD71 spectrum used for comparison (Figures 5 and 6) has been gathered at the HST/CALSPEC spectrophotometric standard stars directory at STScI[5].

The resulting effective temperature and gravity, $T_{\rm eff} = 32,700$ K, log $g = 7.70$, are in agreement with other sources in the literature. For instance, Barstow et al. (2014) estimated $T_{\rm eff} = 32,780$ K and log $g = 7.83$, while Koester et al. (2009) determined $T_{\rm eff} = 32,959$ K and log $g = 7.73$. The stellar distance from the spectral fitting, found to be 48.3 pc, is compatible with the distance of 49 pc found by Gianninas et al. (2011). The model fitting to this HST flux standard reveals a good agreement in overall flux and continuum shape. The model line profiles also reasonably describe the observations, although deeper Balmer line cores were found when both lines and continuum flux are fitted (fig. 6). There is some indication that the observed H$\gamma$ profile is slightly asymmetric, in the sense that it lacks flux in the red wing, while the blue wing is well described by the model profile. Similar discrepancies can also be seen in the fitting of H$\gamma$ and H$\delta$ profiles of a few isolated DAs in Gianninas et al. (2011) and Tremblay et al. (2011). Their origin is not clear and would require further investigation.

---

[4] `classic.sdss.org/dr7/instruments/imager/`

[5] `ftp://ftp.eso.org/pub/stecf/standards/wdstan/`, `http://www.stsci.edu/hst/observatory/cdbs/-astronomical_catalogs.html`

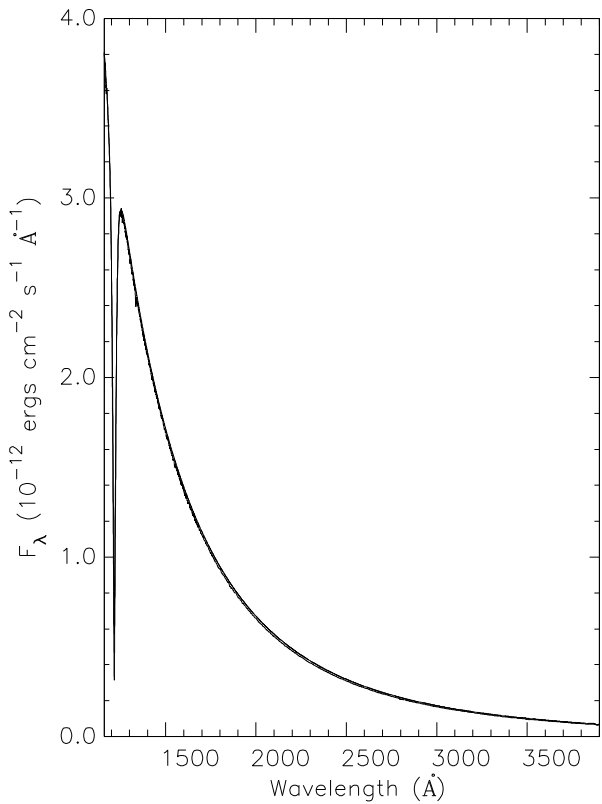

**Figure 5.** Best-fitting non-LTE synthetic spectrum to the observed HST standard WD DA star GD71 between 1200 Å and 4000 Å.

## 5. APPLICATION IN STELLAR POPULATION SYNTHESIS

Spectral stellar libraries are useful in a wide variety of applications, from stellar to extragalactic studies. In extragalactic studies they are, for instance, at the heart of evolutionary stellar population synthesis models, i.e., the modeling of the spectral energy distribution (SED) emitted by evolving stellar populations (e.g. Leitherer et al. 1999; Bruzual & Charlot 2003; Delgado et al. 2005; Coelho et al. 2007; Vazdekis et al. 2015).

Stellar population models should ideally include all stellar evolutionary phases. Nevertheless, due to their low luminosity, WDs have a negligible contribution to the integrated light in the optical region of the spectra (Bruzual & Charlot 2003, G. Bruzual 2017, private communication). In order to investigate their effect on ultraviolet bands, in this section we build stellar population models based on: the stellar evolution predictions from Pietrinferni et al. (2006); Salaris et al. (2010)

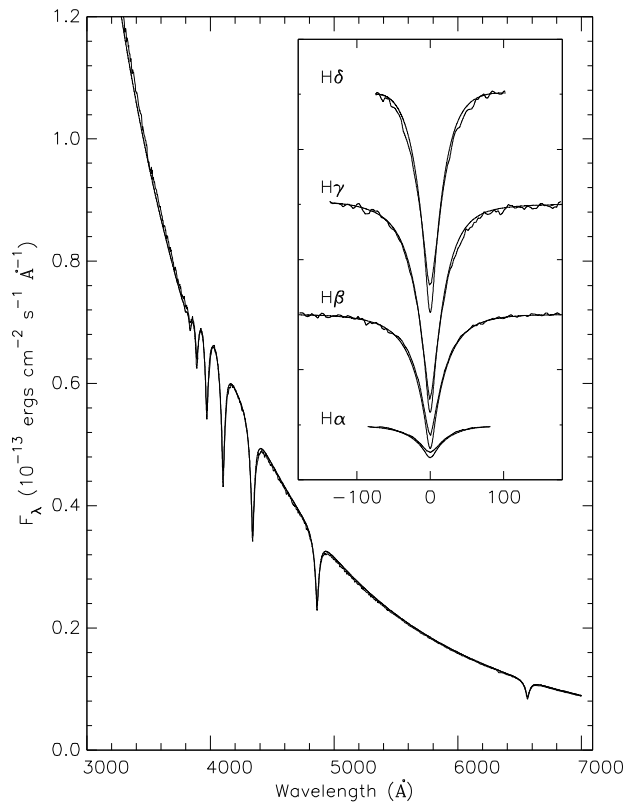

**Figure 6.** Best-fitting of non-LTE models to the observed HST GD71 standard spectrum between 3000 Å and 7000 Å, in the Balmer series range.

(available at BaSTI database[6]; the spectral stellar library of Coelho (2014) to represent stars from the MS to the asymptotic giant branch (AGB); and the grid of the present work to represent WDs. A code in IDL[7] has been written to build stellar population models of simple stellar populations (SSPs) with the ingredients listed above. In short, for each point with an isochrone of age $t$, the spectral stellar grids are interpolated to obtain a spectrum of a given $T_{\rm eff}$, $\log g$ and $\log L/L_\odot$; the resulting stellar fluxes are weighted by the initial mass function of Salpeter (1955) and summed into an integrated spectrum. More details on the isochrone synthesis method of SSP spectra can be found in, e.g., Bruzual A. & Charlot (1993).

In Fig. 7 we show the coverage of atmospheric parameters needed to synthesize stellar populations from 0.5 to 14 Gyr, according to isochrones obtained from BaSTI (gray lines). The isochrones representing stars from the lower MS to the AGB phase are those from Pietrinferni et al. (2006) and correspond to canonical models, α-enhanced composition with metallicity Z = 0.008 and mass-loss η=0.2. Isochrones for the WD cooling sequence correspond to those published in Salaris et al. (2010), adopting He+H envelopes and considering phase separation upon crystallization. The spectral models used in the present synthesis are represented in Fig. 7 as square symbols. Only the DAs effectively used in this application are shown. To represent stars from the MS until the AGB phase, spectral models with a chemical mixture [Fe/H] = -0.8 and [α/Fe] = 0.4 were obtained from Coelho (2014)[8]. This chemical mixture was chosen for this application because it is representative of the metallicity of the thick disk of the Milky Way (e.g. Peng et al. 2013). To represent the cool DAs ($T_{eff}$ < 17000,K), which were not computed in this work but are present in the isochrones, the highest surface gravity spectra from Coelho (2014) for a given temperature were chosen, re-scaled to the luminosity of the DA as given by the isochrone.

---

[6] http://basti.oa-teramo.inaf.it/index.html
[7] Interactive Data Language,
http://www.harrisgeospatial.com/ProductsandSolutions/GeospatialProduc

Figure 8 illustrates the results of the SSP modeling for ages of 1 and 10 Gyr. The SED predictions for the SSPs are given as black lines. The dashed and dotted lines indicate the contributions due to MS–AGB and DAs independently, respectively. MS–AGB stars clearly dominate the flux, as expected, except for the extreme blue at old ages. The contribution from DAs to the integrated SED estimated here may be regarded as a lower limit due to the zero-temperature mass-radius relation adopted, which may underestimate the actual photospheric radius by more than 20% (Wood 1995; Tremblay et al. 2011), depending on the hydrogen layer mass, gravity, and temperature.

In order to identify the ages with a maximum contribution of the DA to the integrated light, SSP models were computed for all available isochrone ages with and without the inclusion of the DAs. Pseudo-magnitudes were measured on the modeled SEDs, in the wavelength ranges as given below, adopting a top-hat throughput:

1. band1: from 1300 to 1800,
2. band2: from 1700 to 2200.

We then compared the evolution of the color C = band1 − band2 predicted for the SSPs. The results, shown in Figure 9, highlight that the maximum contribution of the DA to the integrated light occurs around 8.5 Gyr. This effect is a result of the stellar evolution predictions from Salaris et al. (2010), as seen in Figure

[8] http://specmodels.iag.usp.br

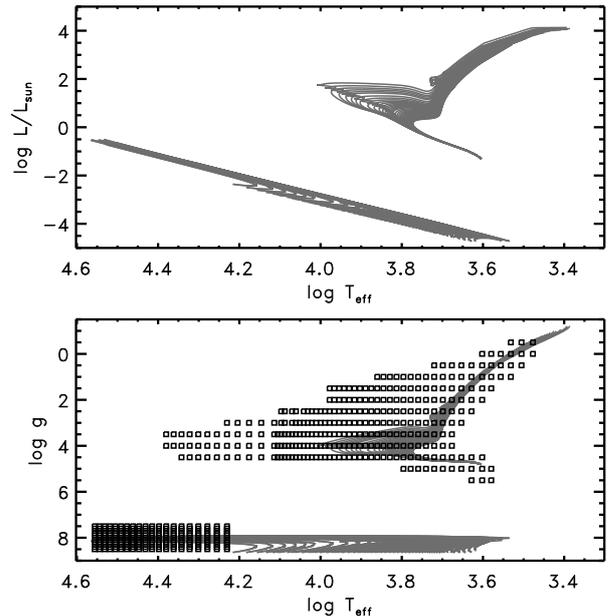

**Figure 7.** Isochrones used in this work (Pietrinferni et al. 2006; Salaris et al. 2010, covering ages from 0.5 to 14 Gyr), are shown as gray lines in the planes $\log T_{eff}$ vs. $\log L/L_\odot$ (upper panel) and $\log T_{eff}$ vs. $\log g$ (lower panel). The square symbols in the lower panel correspond to the a sub-sample of the present DA grid, and the coverage of Coelho (2014), used to represent stars in main-sequence and giant branches.

10: between 7 and 9 Gyr there is not a significant evolution in the locus of MS–AGB stars, but at 8.5 Gyr there is a higher incidence of WDs at $\log T_{eff}$ ∼ 4.5, which then cools down toward higher SSP ages. The contribution from the DA population is most noticeable below 1900 Å where the b–f opacity of Si, C, and Mg dim the spectra of the non-degenerate population. A tiny effect of the WD population is found in the optical: ∼ $10^{-5}$ of the flux around $H_\gamma$ or $H_\beta$, and ∼ $10^{-6}$ around $H_\alpha$, mostly due to the broad lines in DA spectra.

## 6. SUMMARY

In this work we present an extended LTE/non-LTE grid of pure hydrogen model atmospheres and synthetic spectra suitable for hot DA WD studies. The grid covers a range of temperatures, between $17,000\,K \leq T_{eff} \leq 100,000\,K$ with steps of 1000 K and gravities in the range $7.0 \leq \log g \leq 9.5$ with steps of 0.1 dex. All the spectra cover a wide wavelength range, from 900 to 25,000 Å. Modeling the observed spectrum of GD71 results in atmospheric parameters, which agree with the values in the literature. Synthetic HST/ACS, HST/WFC3, Bessel UBVRI and SDSS magnitudes were calculated from the model spectra.

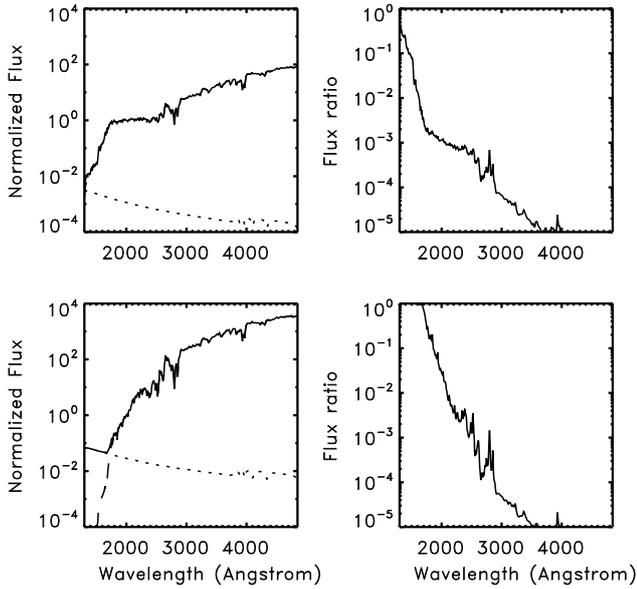

**Figure 8.** SED models for SSPs with ages of 1 Gyr (upper panels) and 10 Gyr (lower panels). In the left panels, the dashed lines are the contribution from stars in main sequence to the AGB and the dotted lines are the contribution from DAs. Total SEDs are indicated in solid lines. Fluxes were normalized at 2000 Å. The right panels show the ratios of the DA contribution over the total SED.

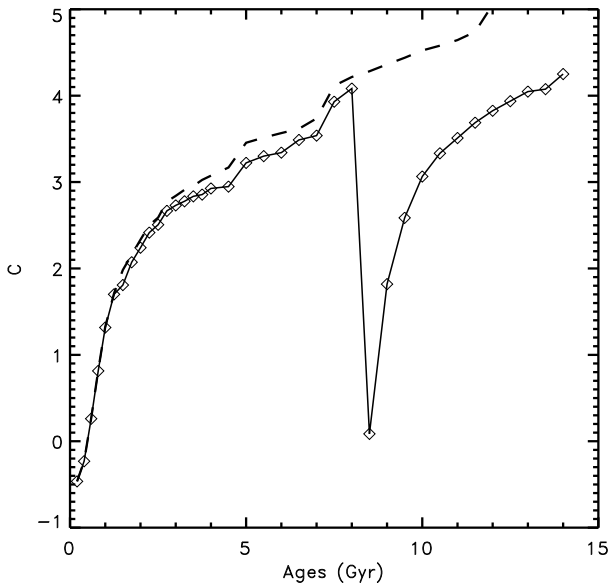

**Figure 9.** Color evolution of SSP models with and without the inclusion of DAs, represented by solid and dashed lines, respectively.

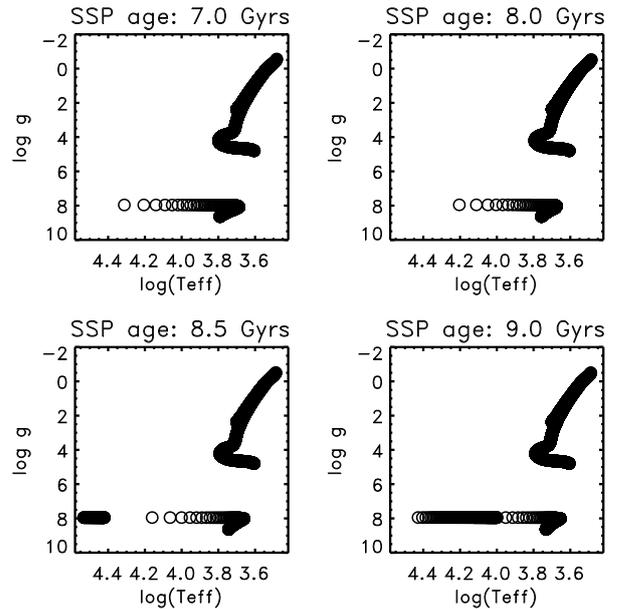

**Figure 10.** Isochrones for 4 different ages as indicated in the panels, from Pietrinferni et al. (2006) and Salaris et al. (2010).

We also employed this grid in the modeling of the integrated properties of an SSP, using the technique of stellar population isochrone synthesis. We reproduce previous results that show that the effect of WDs on the integrated light is negligible in the optical bands, even for high S/N observations, but we highlight that our modeling suggests that their effect might be measurable in ultraviolet bands for populations older than $\sim 8$ Gyr. This is the first application of an ongoing investigation of the effect of low-luminosity hot components lurking in the integrated light of old stellar populations. These simulations will be expanded in forthcoming work, considering more realistic mass-radius relations for WDs, different IMFs and the contribution from sub-dwarfs and interacting binary systems.

M.P.D. acknowledges the support from CNPq under grant #305657.

*Software:* TLUSTY, SYNSPEC, ROTIN